\documentclass[twocolumn, pra, showpacs,superscriptaddress,,floatfix]{revtex4}
\usepackage{epsfig}
\usepackage{amsmath}

\begin{document}

\title{Ground state properties of one-dimensional ultracold Bose gases in a hard-wall
trap}
\author{Yajiang Hao}
\affiliation{Department of Physics and Institute of Theoretical
Physics, Shanxi University, Taiyuan 030006, P. R. China}
\affiliation {Beijing National Laboratory for Condensed Matter
Physics, Institute of Physics, Chinese Academy of Sciences,
Beijing 100080, P. R. China}
\author{Yunbo Zhang}
\affiliation{Department of Physics and Institute of Theoretical
Physics, Shanxi University, Taiyuan 030006, P. R. China}
\author{J. Q. Liang}
\affiliation{Department of Physics and Institute of Theoretical
Physics, Shanxi University, Taiyuan 030006, P. R. China}
\author{Shu Chen}
\email{schen@aphy.iphy.ac.cn}
\affiliation{Beijing National
Laboratory for Condensed Matter Physics, Institute of Physics,
Chinese Academy of Sciences, Beijing 100080, P. R. China}

\date{\today}

\begin{abstract}
We investigate the ground state of the system of $N$ bosons
enclosed in a hard-wall trap interacting via a repulsive or
attractive $\delta$-function potential. Based on the Bethe ansatz
method, the explicit ground state wave function is derived and the
corresponding Bethe ansatz equations are solved numerically for
the full physical regime from the Tonks limit to the strongly
attractive limit. It is shown that the solution takes different
form in different regime. We also evaluate the one body density
matrix and second-order correlation function of the ground state
for finite systems. In the Tonks limit the density profiles
display the Fermi-like behavior, while in the strongly attractive
limit the Bosons form a bound state of $N$ atoms corresponding to
the $N$-string solution. The density profiles show the continuous
crossover behavior in the entire regime. Further the correlation
function indicates that the Bose atoms bunch closer as the
interaction constant decreases.
\end{abstract}
 \maketitle



\narrowtext

The physics of one-dimensional (1D) cold atoms has recently
attracted a great amount of attention due to tremendous
experimental progress in the realization of trapped 1D cold atom
systems \cite {gorlitz,Paredes,Toshiya,esslinger,Tolra}. A 1D
quantum gas is obtained by tightly confining the particle motion
in two directions to zero point oscillations \cite{Ketterler}. As
the radial degrees of freedom is frozen, the quantum gas is
effectively described by a 1D model along the longitudinal
direction \cite{Olshanii}. Experimentally, a 1D Bose gas can be
realized either by means of anisotropic magnetic trap or
two-dimensional optical lattice potentials. In parallel, the wide
exploitation of the Feshbach resonance to control the scattering
length of the atoms allowed experimental access to the full regime
of interactions both from weakly interacting limit to strongly
interacting limit and from the repulsive regime to the attractive
regime by simply tuning a magnetic field. Very recently, several
groups have reported the observation of a 1D Tonks-Girardeau (TG)
gas \cite{Paredes,Toshiya,Tolra}, which provides a textbook
example where atom-atom interactions play a critical role and the
mean-field theory fails to obtain reasonable results
\cite{Girardeau2}. Theoretically, the effect of dimensionality and
correlation effect of bosonic system have been investigated
extensively \cite
{Olshanii,Olshanii2,Petrov,Dunjko,Luxat,Pedri,Kolomeisky,Chen,Kunal,ohberg,Girardeau1}
and is being paid more and more attention.

It is well known that there exist some exactly solved 1D
interacting models \cite{Lieb,Yang,Takahashi,Korepin,McGuire},
however, are not directly applicable to the system trapped in an
external harmonic potential and the Gross-Pitaviskii (GP) theory
is widely adopted in dealing with the system of Bose-Einstein
condensations (BECs) with weak interactions. In spite of its great
success in accounting for the basic experimental observations, the
GP equation is essentially based on a mean-field approximation and
suffers from various shortcomings, especially when applied to the
strongly correlated systems. Some recent works have shown the
limitations of GP theory in the density distribution of 1D trapped
gas in strongly interacting regime\cite{Kolomeisky}, the
overestimation of interference\cite{Chen}, and the instability of
GP equation with attractive interactions\cite{Holland}. For a
trapped system, the attractive interactions may compensate the
kinetic energy of the condensate and lead to a stable soliton
solution. Recently, the 1D GP equation under box boundary
condition was solved analytically for both repulsive and
attractive condensates\cite{Carr}. On the other hand, the
experimental progress in building up square well trap\cite{Hansel}
and optical box trap \cite{Raizen}, gives rise to the hope to
directly study the physics in the textbook geometry of a
``particle in a box''. It is not clear whether the solution of GP
equation is good enough to describe the interacting Bose gas in
the hard-wall trap, especially for the attractive case and under
the strongly interacting limit. Fortunately, in a hard-wall trap,
the corresponding interacting model is integrable and its exact
solution has been obtained with the Bethe ansatz method for the
repulsive interaction in a seminal paper by Gaudin\cite{Gaudin}.
So far, there has been a growing interest in the exactly solved
models in the hard-wall trap\cite{Guan,Guan2,SJGu,Muga}. In spite
of the long history of the integrable Bose interacting model
(Lieb-Liniger model), the case with attractive interaction draws
less attention\cite{McGuire,Muga98,Sakmann} and most of the
studies have focused on the ground state energy for the periodic
boundary systems.

In this paper, we investigate the density distribution of the ground state
of the 1D Bose gases in an infinite deep square potential well. Different
from the system with periodic boundary condition where the density
distribution is a constant and is irrelevant to the strength of interaction,
the density profile for a trapped bosonic system is shown to be sensitive to
its interaction. The dependence of one body density matrix on the
interaction between atoms and the effects of finite size are studied for
both the repulsive and attractive interactions. The wave function of ground
state is obtained by numerically solving the set of Bethe ansatz equations.

\begin{figure}[tbp]
\includegraphics[width=3.2in]{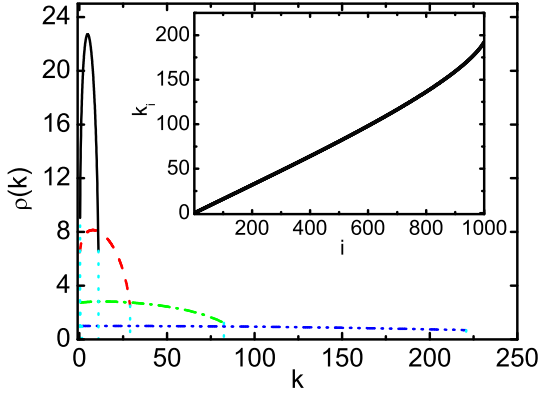}
\caption{(color online) The density of state in quasi-momentum $k$
space for the ground state for $N=200$ and $c=0.1$ (solid line),
$c=1$ (dashed lines), $c=10$ (dash dot lines), $c=100$ (dash dot
dot lines). Inset: The numerical solutions of the Bethe ansatz
equations for $N=1000$ and $c=10$.} \label{fig1}
\end{figure}
We consider $N$ particles with the $\delta $ interaction in one dimensional
box with length $L$. The Schr\"{o}dinger equation can be formulated as (the
natural unit is used)
\[
\left[ -\sum_{i=1}^N\frac{\partial ^2}{\partial x_i^2}+2c\sum_{1\leq i<j\leq
N}\delta \left( x_i-x_j\right) \right] \Psi =E\Psi ,
\]
where $2c$ is the interaction strength between atoms related with the $s$
-wave scattering length \cite{Olshanii,Olshanii2}, which can be tuned from $%
-\infty $ to $+\infty $ by the Feshbach resonance or confinement induced
resonance. When the interaction between atoms is repulsive, $c>0$, and in
the attractive case $c<0$. The important parameter characterizing different
physical regimes of 1D quantum gas is $\gamma =c\rho ^{-1}$, where $\rho =N/L
$. We will study the full physical regimes $-\infty <c<\infty $. The
wavefunction can be written as
\begin{eqnarray*}
\Psi \left( x_1,\cdots ,x_N\right) &=&\sum_P\varphi \left(
x_{p_1},x_{p_2},\cdots ,x_{p_N}\right) \\
&&\times \theta \left( x_{p_1}<x_{p_2}\cdots <x_{p_N}\right) ,
\end{eqnarray*}
in which
\[
\theta \left( x_{p_1}<\cdots <x_{p_N}\right)  = \theta \left(
x_{p_N}-x_{p_{N-1}}\right) \cdots \theta \left( x_{p_2}-x_{p_1}\right),
\]
$p_1,p_2,\cdots ,p_N$ is one of the permutations of $1,\cdots ,N$,
and $ \sum_P $ is the sum of all permutations. $\theta(x-y)$ is
the step function. The wavefunction $\varphi \left(
x_{p_1},x_{p_2},\cdots ,x_{p_N}\right) $ could be obtained by the
permutation of $\varphi \left( x_1,\cdots ,x_N\right) $ according
to the symmetry condition of the Boson wave function. It turns out
that the original problem is equivalent to solving the equation
\begin{eqnarray}
&&\left[ -\sum_{i=1}^N\frac{\partial ^2}{\partial x_i^2}+2c\sum_{1\leq
i<j\leq N}\delta \left( x_i-x_j\right) \right] \varphi \left( x_1,\cdots
,x_N\right)  \nonumber \\
&=&E\varphi \left( x_1,\cdots ,x_N\right)  \label{sch}
\end{eqnarray}
in the region $0\leq x_1\leq x_2\leq \cdots \leq x_N\leq L$. We take the
wave function $\varphi \left( x_1,\cdots ,x_N\right) $ as the Bethe ansatz
type
\begin{equation}
\varphi \left( x_1,x_2,\cdots ,x_N\right) =\sum_{P,r_1,\ldots
,r_N}\left[ A_{P}\exp \left( i\sum_jr_jk_{p_j}x_j\right) \right] ,
\label{wavfunc}
\end{equation}
where $r_j=\pm $ indicates that the particles move toward the right or the
left. Substituting eq.(\ref{wavfunc}) into eq.(\ref{sch}) and using the open
boundary conditions
\[
\varphi \left( 0,x_2,\cdots ,x_N\right) =\varphi \left( x_1,x_2,\cdots
,L\right) =0,
\]
we have the Bethe ansatz equations
\begin{equation}
\exp \left( i2k_jL\right) =\prod_{l=1\left( \neq j\right)
}^N\frac{
ik_l+ik_j-c}{ik_l+ik_j+c}\frac{ik_l-ik_j+c}{ik_l-ik_j-c},
\label{BAE1}
\end{equation}
with $j=1,2,\cdots,N$. The energy eigenvalue is
$E=\sum_{j=1}^Nk_j^2$ and the total momentum is $
k=\sum_{j=1}^Nk_j$.

Taking the logarithm of Bethe ansatz equations, we have
\begin{equation}
k_jL=n_j\pi +\sum_{l=1\left( \neq j\right) }^N\left( \arctan \frac
c{k_j+k_l}+\arctan \frac c{k_j-k_l}\right)  \label{bae}
\end{equation}
with $j=1,\cdots,N$. Here $\left\{ n_j\right\} $ is a set of
integer which determines an eigenstate and for the ground state
$n_j=1$ ($1\leq j\leq N$). Alternatively, eq. (\ref{bae}) has the
form of
\begin{equation}
k_jL=n_j^{\prime }\pi -\sum_{l=1\left( \neq j\right) }^N\left( \arctan \frac{
k_j+k_l}c+\arctan \frac{k_j-k_l}c\right) .  \label{baeb}
\end{equation}
The above two equations are consistent but the choice of $\left\{
n_j^{\prime }\right\} $ is different from that of $\left\{ n_j
\right\} $. For the latter case, we should have $n_j^{\prime }=j$
($1\leq j\leq N$ ) for the ground state. The $\left\{ k_j\right\}
$, and thus the wave function can be decided by the set of
transcendental equations eqs.(\ref{bae}). In general, the
eqs.(\ref{baeb}) are used in the literature \cite
{Takahashi,Korepin} for the model with repulsive interaction,
because it is more convenient to extend it to deal with the
problem of thermodynamics. For $c>0$, the set of $k_i$ is unique
and real, therefore the density of state in $k$ space $\rho(k)$
can be formulated as \cite{Yang}
\begin{equation}
L\rho \left( \frac{k_j+k_{j+1}}2\right) =\frac 1{k_{j+1}-k_j}.  \label{d_k}
\end{equation}
However, such a definition does not make sense when $c<0$. In the
later, through an example with $N=3$, we will show that in the
attractive case the solution is not unique corresponding to a
given set of $\left\{ n_j \right\} $ and the ground state is
decided by comparing the energy eigenvalues of different
solutions.

With some algebraic calculation, the wave function has the following
explicit form
\begin{eqnarray*}
&&\varphi \left( x_1,x_2,\cdots ,x_N\right) \\
&=&\sum_PA_p\in _p\exp \left[ i\left( \sum_{l<j}^{N-1}\omega
_{p_jp_l}\right) \right] \exp \left( ik_{p_N}L\right) \sin \left(
k_{p_1}x_1\right) \\
&&\times \prod_{1<j<N}\sin \left( k_{p_j}x_j-\sum_{l<j}\omega
_{p_lp_j}\right) \sin \left( k_{p_N}\left( L-x_N\right) \right)
\end{eqnarray*}
with
\[
\omega _{ab}=\arctan \frac c{k_b-k_a}+\arctan \frac c{k_b+k_a}
\]
and
\[
A_{p_1p_2...p_N}=\prod_{j<l}^N\left( ik_{p_l}-ik_{p_j}+c\right) \left(
ik_{p_l}+ik_{p_j}+c\right) .
\]
Here $\in _p$ denotes a $+(-)$ sign factor associated with even
(odd) permutations. In terms of the ground state wave function $\Psi
\left( x_1,\cdots ,x_N\right) $, the one body density matrix is
defined as
\[
\rho(x) =\frac{N\int_0^Ldx_2\cdots dx_N\left|\Psi
\left(x,x_2,x_3,\cdots,x_N\right)\right|^2}{\int_0^Ldx_1\cdots
dx_N\left| \Psi \left( x_1,x_2,\cdots ,x_N\right) \right| ^2}.
\]

In the following calculation, we will let $L=1$ through the paper.
Firstly, the repulsive interaction case is considered. The density
of ground state in quasi-momentum $k$ space for different
interaction constants is plotted in Fig. 1. It is shown that the
density is suppressed for $k$ close to zero because of the
confinement by the infinite-depth well. The numerical results of
eqs. (\ref{bae}) are given for $N=1000$ and $c=10$ in the inset.
Bethe ansatz equations uniquely decide the value of $k_i$, the
wave function of the ground state and the one body density matrix
of the system. In Fig. 2, we show the one body density matrix of
the ground state for $N=4$ for different interaction constant.
When there is no interaction between atoms ($c=0$), all of the
atoms have the same quasi momentum, $k_j=\pi $ ($1\leq j\leq N$),
corresponding to condensation of the ideal Bose gas. The one body
density matrix is the sum of the density of $N$ independent
bosonic atoms lying in the ground state. As the interaction
increases further, the half-width of the density of the system
becomes larger and larger with increasing interaction. At $c=10$,
the density distribution shows already Fermion-like behavior. As
the further increase of interaction the Bosons display the same
density profile as that of $N$ noninteracting spinless Fermions
and we have $k_j\simeq j\pi $ ($1\leq j\leq N$) at $c\sim1000$.

\begin{figure}[tbp]
\includegraphics[width=3.2in]{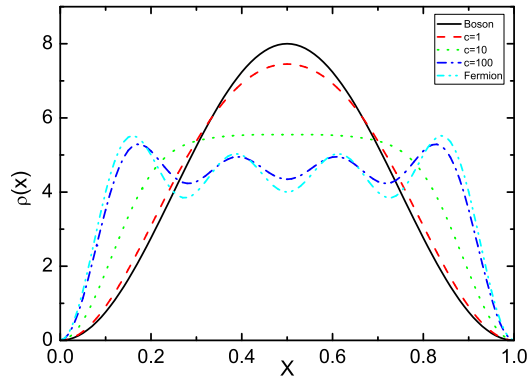}
\caption{(color online) The density profiles $\rho(x)$ of the
ground state for different repulsive interaction constants and
$N=4$.} \label{fig2}
\end{figure}

\begin{figure}[tbp]
\includegraphics[width=3.0in]{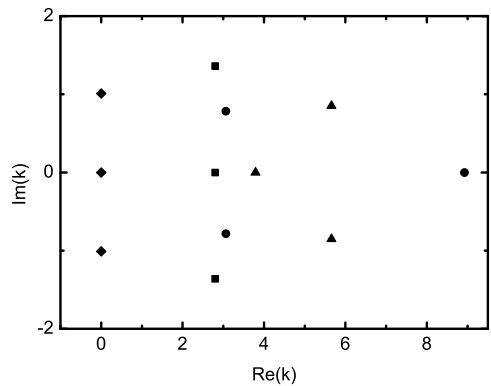}
\caption{Three sets of solution (squares,triangles,circles) for 3
Bosons in a hard-wall trap with the attractive interaction $c=-1$.
As comparison, diamonds correspond to the ground state solution of
a periodic 3-Boson system on a ring of length $L=1$ for $c=-1$. }
\label{fig3}
\end{figure}
In the case of attractive interaction, $c<0$, eqs. (\ref{bae})
have solutions with complex quasi-momentum. The case for
periodical boundary condition has been investigated in reference
\cite{Takahashi,Muga98,Sakmann}. Due to the zero-point energy for
a confined system, the solution corresponding to the ground state
is not purely imaginary like that for the Bethe ansatz equations
with periodical boundary condition. For the two body problem, the
complex solutions of the Bethe ansatz equations take the form of
the two-string solution
\begin{eqnarray*}
k_1 &=&\alpha -i\Lambda , \\
k_2 &=&\alpha +i\Lambda ,
\end{eqnarray*}
where $\alpha$ and $\Lambda $ are real and $2\alpha$ is the momentum
of the system. In  general, by an $N$-string solution we mean a
solution where $N$  momenta $k_i$ possess the same real part. Then
the eqs. (\ref{bae}) have the following form
\begin{eqnarray}
\alpha L &=& \frac {n_1 + n_2}{2} \pi + \arctan {\frac {c}{2\alpha},}  \nonumber \\
\Lambda L &=&\frac 12\ln \frac{2\Lambda -c}{2\Lambda +c},
\end{eqnarray}
where $n_1=n_2=1$ for the ground state. The corresponding energy
of the system is $E=2 \alpha^2-2 \Lambda^2$. In terms of the
momentum $k=2\alpha$, the energy can be represented as $E=k^2/2 -
2 \Lambda^2$, which implies that the state corresponding to a
two-string solution can be regarded as a bound state of two atoms
with a binding energy $-2 \Lambda^2$ and a doubled mass. For
convenience, we also refer to the bound state corresponding to a
two-string solution as a dimer state in the following text.

For the three-body problem, we assume that the solution has the
following form: $ k_1=\alpha -i\Lambda ,$ $k_2=\alpha +i\Lambda $
and $k_3=\beta ,$ therefore the Bethe ansatz equations can be
represented with $\alpha $, $\beta $ and $ \Lambda $. For the case
with periodic boundary condition, there exists a 3-string solution
corresponding to $\alpha=\beta $, i.e. the trimer state (the bound
state of three particles) \cite{Takahashi,Muga98}.  But we can
prove that no 3-string solution could be formed in a box for
finite attractive interaction (see the appendix for details). That
means the open boundary condition or a confinement tends to
prevent the formation of a trimer state. We show the solutions of
three-body problem for periodical boundary condition (Diamonds)
and open boundary condition (Squares,Triangles and Circles
denoting three sets of different solutions) for $c=-1$ in Fig. 3.
It should be noticed that the set of solution denoted with squares
is not an exact 3-string solution although $\alpha \approx \beta$,
which is a dimer plus a single atom. By comparing $E$, the
trimer-like solution (squares) has the lowest energy eigenvalue
and thus it represents the ground state of the system.
\begin{figure}[tbp]
\includegraphics[width=3.5in]{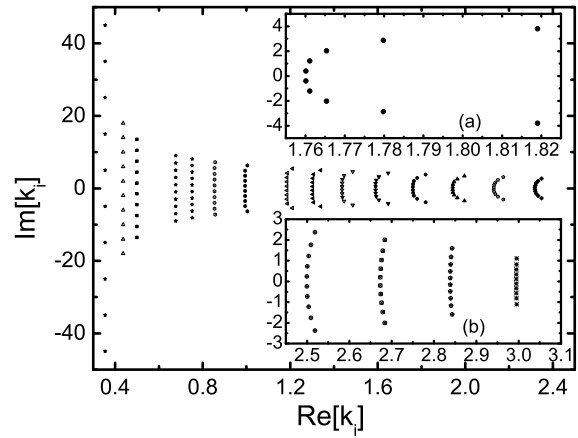}
\caption{The solutions to the Bethe ansatz equations for $N=10$ with
different attractive interaction constants. $c$ takes the value of $-10$, $%
-4.0$, $-3.0$, $-2.0$, $-1.8$, $-1.6$, $-1.4$, $-1.2$, $-1.1$, $%
-1.0 $, $-0.9$, $-0.8$, $-0.7$, $-0.6$, $-0.5$ (from left to
right). Inset: (a) $c=-0.8$; (b) $c$ takes the values of $-0.4$,
$-0.3$, $-0.2$, $-0.1$ (from left to right).}\label{fig4}
\end{figure}

For the system with four atoms, the general complex solutions can
be assumed as $k_1=\alpha-i\Lambda_1,$ $k_2=\alpha +i\Lambda_1,$
$k_3=\beta -i\Lambda_2$ and $k_4=\beta +i\Lambda _2$. By solving
the Bethe ansatz equations numerically, it turns out that the
ground state corresponds to a string solution of length four
($\alpha=\beta$ but $\Lambda_1\neq \Lambda_2$) for the large
attractive interaction, whereas the ground state is a scattering
state of two dimers for the weak attractive interaction. For $N>4$
the numerical results show that generally the ground state of the
system in the weak attractive regime is a scattering state of
$N/2$ dimers ($N$ is even) or $(N-1)/2$ dimers plus a single atom
($N$ is odd). The state of $N/2$ dimers is given in terms of $N/2$
two-string solutions. As the attractive interaction increases
further, the ground state smoothly evolves into an intermediate
regime characterized by an $(N-M)$-string solution plus $M/2$
two-string solutions with $2\leq M \leq N$, and finally it falls
into the strong coupling regime corresponding to the $N$-string
solution. To give an explicit example, we display the ground state
solutions to the eqs.(\ref{bae}) for $N=10$ with different finite
attractive interactions in Fig. 4. Based on the numerical
solutions of the Bethe ansatz equations, we find that $5$ dimers
form the ground state of the system when the attractive
interaction is weak ($-1.42<c<0$). While in the limit of strong
attractive interaction($c\leq-3.02$), the ground state solution is
a 10-string solution. In the intermediate regime ($-3.02<
c\leq-1.42$), the ground state solution is composed of a 4-string
solution plus three 2-string solutions for $-1.72 <c\leq-1.42$, a
6-string solution plus two 2-string solutions for $-1.88< c\leq
-1.72$, and an 8-string solution plus a 2-string solution for
$-3.02 < c \leq -1.88$.

For the periodic system, the $N$-string ansatz is represented as
\[
k_j=\alpha + i(N+1-2j)(c/2), ~~~~j=1,\cdots,N.
\]
The N-string ansatz is generally believed to be the solution of the
Bethe ansatz equations for $c<0$ in the limit $L \rightarrow
\infty$\cite{Takahashi}. For the Bose gases in a finite-size
hard-wall trap, we assume that the $N$-string  solution takes the
following form
\begin{eqnarray}
k_j=\alpha + i(N+1-2j)(c/2+\delta_j), ~~~~j=1,\cdots,N,
\label{infin}
\end{eqnarray}
where ${\delta_j}$ is a set of small numbers and $\delta_j
\rightarrow 0$ as  $c \rightarrow -\infty$. Our numerical
solutions to eqs.(\ref{bae}) indicate that the solutions are
precisely fitted by the $N$-string ansatz of eqs.(\ref{infin}) in
the limit of large attractive interaction. For instance, for the
system with ten atoms the solution takes the values of
$k_j=0.498075-i(11-2j)(5+\delta_j)$ with $|\delta_j|<10^{-8}$ for
$c=-3.02$ and $k_j=0.354213-i(11-2j)(5+\delta_j)$ with
$|\delta_j|<10^{-13}$ for $c=-10$. In the strong attractive limit,
we can determine $\alpha$ analytically. With eq. (\ref{infin}) and
the original Bethe ansatz equations the total momentum can be
formulated as
\[
\exp \left( i2kL\right) =\prod_{1\leq j<N}\left[ \frac{2\alpha
+i(j c)}{2\alpha -i (j c)}\right] ^2,
\]
where $k=N\alpha$ and we have taken $\delta_i=0$. Taking the
logarithm to the above equation, we have
\begin{eqnarray}
kL=n\pi +\sum_{1\leq j<N}2\arctan \frac{j c }{2\alpha },
\end{eqnarray}
where $n=N$ for the ground state. It is clear that the momentum
$k$ is $\pi$ in the limit of $c\rightarrow-\infty$ and therefore
$\alpha = \pi/N$ in the strong attractive limit.

In Fig. 5 we plot the one body density matrix of the ground state
for $N=2$ (a) and $N=4$ (b). It is shown that as the attractive
interaction increases, the central density of the system becomes
large firstly and then less. In the limit of $c\rightarrow -\infty $
($c\sim -1000$ for $N=2$), the density profile matches the case of
$c=0$. The matching between them can be explained as a compounded
particle with mass $Nm$ located in its ground state, whose density
distribution has the form of $\sin^2x$ in the strong attractive
limit. When the solution of Bethe ansatz equations take the form of
eq.(\ref{infin}), the ground state energy can be represented as
$E=k^2/N - N(N^2-1)c^2/12$, which implies that the state
corresponding to the $N$-string solution can be regarded as a bound
state of $N$ atoms with a binding energy $\propto - c^2$ and a
 mass $N$m.  It is also interesting to study the second order correlation
 function
\[
g_2(x,c)=\frac{N(N-1)\int_0^Ldx_3\cdots dx_N\left|\Psi
\left(x,x,x_3,\cdots,x_N\right)\right|^2}{\int_0^Ldx_1\cdots
dx_N\left|\Psi \left(x_1,x_2,\cdots,x_N\right)\right|^2}.
\]
In Fig. 6 we show the dependence of second-order correlation
function upon the interaction. It indicates that the atoms tend to
cluster together more easily for the attractive interaction and
the atoms bunch closer as the interaction becomes stronger. For
the repulsive interactions, the atoms avoid each other and the
atom-bunching reduces and vanishes finally for increasing
interactions, which is similar to the case for the periodical
boundary condition \cite{Shlyapnikov}.
\begin{figure}[tbp]
\includegraphics[width=3.7in]{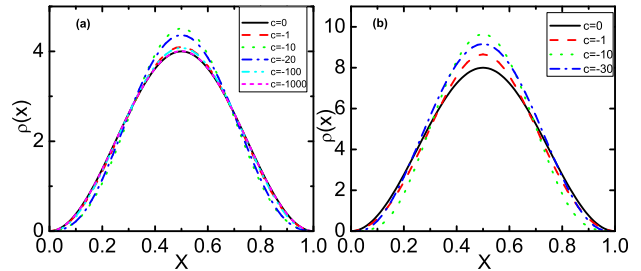}
\caption{(color online) The density profiles $\rho(x)$ of the
ground state for different attractive interaction constants. (a)
$N=2$ and (b) $N=4$.} \label{fig5}
\end{figure}
\begin{figure}[tbp]
\includegraphics[width=3.2in]{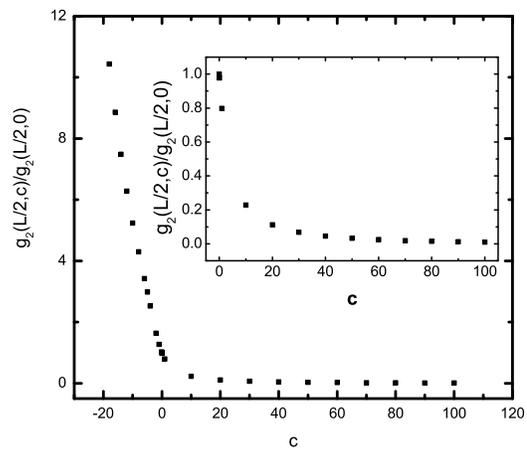}
\caption{The second-order correlation function versus $c$ for $N=4$.
Inset: The second-order correlation function for repulsive
interaction with $N=4$.} \label{fig6}
\end{figure}

In conclusion, by numerically solving the Bethe ansatz equations
we investigate the ground state properties and obtain the density
distribution function and the second-order correlation function of
the 1D Bose gases in a box of finite length $L$ in all the
physical regimes ($-\infty<c<+\infty$). In the limit of
$c\rightarrow \infty$ the Bose gas shows similar behavior as that
of noninteracting spinless Fermions, while in the limit of
$c\rightarrow -\infty$ the Bose gas behaves as a compounded
particle with mass $Nm$. In the case of weakly attractive
interaction the ground state is composed of $N/2$ dimers ($N$ is
even) or $(N-1)/2$ dimers plus a single atom ($N$ is odd). The
second-order correlation function indicates that the atoms bunch
closer as the interacting constant $c$ decreases. Our results can
cover the whole parameter regime beyond the mean field theory and
display the continuous crossover behavior from the Tonks limit to
the strongly attractive limit. Especially, the crossover behavior
of the ground state for the case of attractive interactions is
discussed in detail. Hopefully, our results based on the exact
solution can provide a clear picture of the wavefunction for the
BECs with attractive interaction in a trap and help us to gain
some intuitive insight on the collapse of BEC.

S.C. would like to acknowledge helpful discussions with Y. Wang,
W. D. Li and S. J. Gu. He also thanks the Chinese Academy of
Sciences for financial support. This work is supported in part by
NSF of China under Grant No. 10574150. Y.Z. thanks the hospitality
of Prof. K.-A. Suominen and the Quantum Optics group for their
hospitality at University of Turku, where some of this work was
done. He is also supported by Shanxi Province Youth Science
Foundation under grant No. 20051001.
\appendix
\section{The string solution for the
attractive three-atom system}
In this appendix, we study the three-particle system in detail. We
firstly assume the 3-string solution for 3-atom system in the form
of $ k_1=\alpha -i\Lambda ,$ $k_2=\alpha +i\Lambda $ and
$k_3=\alpha .$ In terms of $\alpha $ and $\Lambda ,$ then the
Bethe ansatz equations take the following forms
\begin{eqnarray}
&&\exp \left[ \left( i2\alpha +2\Lambda \right) L\right]   \nonumber \\
&=&\frac{i2\alpha +\Lambda -c}{i2\alpha +\Lambda +c}\frac{i2\alpha -c}{%
i2\alpha +c}\frac{\Lambda -c}{\Lambda +c}\frac{2\Lambda -c}{2\Lambda +c}, \\
&&\exp \left[ \left( i2\alpha -2\Lambda \right) L\right]   \nonumber \\
&=&\frac{i2\alpha -c}{i2\alpha +c}\frac{i2\alpha -\Lambda
-c}{i2\alpha
-\Lambda +c}\frac{2\Lambda +c}{2\Lambda -c}\frac{\Lambda +c}{\Lambda -c}, \\
&&\exp \left[ i2\alpha L\right]   \nonumber \\
&=&\frac{i2\alpha +\Lambda -c}{i2\alpha +\Lambda +c}\frac{i2\alpha
-\Lambda -c}{i2\alpha -\Lambda +c},
\end{eqnarray}
where we have used $\frac{\Lambda +c}{\Lambda -c}\frac{\Lambda -c}{\Lambda +c%
}=1$ in the right of the third equation. Substituting eq. (A3)
into (A1) and (A2), we then get
\begin{eqnarray*}
\exp \left[ 2\Lambda L\right] \frac{i2\alpha -\Lambda -c}{i2\alpha
-\Lambda
+c} &=&\frac{i2\alpha -c}{i2\alpha +c}\frac{\Lambda -c}{\Lambda +c}\frac{%
2\Lambda -c}{2\Lambda +c}, \\
\exp \left[ -2\Lambda L\right] \frac{i2\alpha +\Lambda -c}{i2\alpha
+\Lambda
+c} &=&\frac{i2\alpha -c}{i2\alpha +c}\frac{2\Lambda +c}{2\Lambda -c}\frac{%
\Lambda +c}{\Lambda -c}.
\end{eqnarray*}
The above two equations can be rewritten as
\begin{eqnarray*}
&&\exp \left[ 2\Lambda L\right] \sqrt{\frac{4\alpha ^2+\left(
\Lambda+c\right) ^2}{4\alpha ^2+\left( \Lambda -c\right) ^2}} \\
&&\times \exp \left[ i\left( \arctan \frac{\Lambda +c}{2\alpha
}-\arctan
\frac{\Lambda -c}{2\alpha }\right) \right]  \\
&=&\exp \left[ 2i\arctan \frac c{2\alpha }\right] \frac{\Lambda
-c}{\Lambda
+c}\frac{2\Lambda -c}{2\Lambda +c}, \\
&&\exp \left[ -2\Lambda L\right] \sqrt{\frac{4\alpha ^2+\left(
\Lambda
-c\right) ^2}{4\alpha ^2+\left( \Lambda +c\right) ^2}} \\
&&\times \exp \left[ i\left( \arctan \frac{\Lambda +c}{2\alpha
}-\arctan
\frac{\Lambda -c}{2\alpha }\right) \right]  \\
&=&\exp \left[ 2i\arctan \frac c{2\alpha }\right] \frac{2\Lambda +c}{%
2\Lambda -c}\frac{\Lambda +c}{\Lambda -c}.
\end{eqnarray*}
Comparing the module or imaginary part of the left and right side of
these two equations, we can get three equations corresponding to
Eqs. (A1), (A2) and (A3)
\begin{eqnarray}
\exp \left[ 2\Lambda L\right]=\sqrt{\frac{4\alpha ^2+\left( \Lambda
-c\right) ^2}{4\alpha ^2+\left( \Lambda +c\right) ^2}}\frac{\Lambda
-c}{\Lambda +c}\frac{2\Lambda -c}{ 2\Lambda +c},
\end{eqnarray}
\begin{eqnarray}
&&\exp \left[ 2i\arctan \frac c{2\alpha }\right] \nonumber \\
&=&\exp \left[ i\left( \arctan \frac{\Lambda +c}{2\alpha }-\arctan \frac{%
\Lambda -c}{2\alpha }\right) \right] , \\
&&\exp \left[ i2\alpha L\right]   \nonumber \\
&=&\exp \left[ 2i\left( \arctan \frac{\Lambda +c}{2\alpha }-\arctan \frac{%
\Lambda -c}{2\alpha }\right) \right] .
\end{eqnarray}
Observing that there are two unknown parameters $\alpha $ and
$\Lambda$ but they need fulfill three equations, there are no
solutions to the above equations for an arbitrary $c$. This means
that the three-string assumption is not a solution to the Bethe
ansatz equations. Also, we have solved the above equations
numerically, but no solutions are found.

Next we assume that the solution for a 3-atom system has the
following form $ k_1=\alpha -i\Lambda ,$ $k_2=\alpha +i\Lambda $
and $k_3=\beta $ with $\Lambda>0$. With similar procedure as
above, the Bethe ansatz equations can be represented as
\begin{eqnarray}
\left( 2\alpha +\beta \right) L &=&\left( 2n+n^{\prime }\right) \pi
+2\left(
\arctan \frac{\Lambda +c}{\alpha +\beta }\right. \nonumber \\
&&\left. -\arctan \frac{\Lambda -c}{\alpha +\beta }+\arctan \frac
c{2\alpha } \right), \label{3body1}  \\
\left( 2\alpha -\beta \right) L &=&\left( 2n-n^{\prime }\right) \pi
+2\left(
\arctan \frac{\Lambda +c}{\alpha -\beta }\right.  \nonumber \\
&&\left. -\arctan \frac{\Lambda -c}{\alpha -\beta }+\arctan \frac c{2\alpha }%
\right), \label{3body2}  \\
\exp \left[ 2\Lambda L\right]  &=&\frac{\left[ \left( \alpha
+\beta \right) ^2+\left( \Lambda -c\right) ^2\right]
^{1/2}}{\left[ \left( \alpha +\beta \right) ^2+\left( \Lambda
+c\right) ^2\right] ^{1/2}} \nonumber \\
&&\frac{\left[ \left( \alpha -\beta \right) ^2+\left( \Lambda
-c\right) ^2\right] ^{1/2}}{\left[ \left( \alpha -\beta \right)
^2+\left( \Lambda +c\right) ^2\right] ^{1/2}}\frac{2\Lambda
-c}{2\Lambda +c}.
\end{eqnarray}
where $n$ and $n^{\prime }$ are integers which can be determined
in the limit $c\rightarrow 0$. Here $\left( 2\alpha +\beta \right)
$ corresponds to the total momentum which is conserved in the
periodic boundary condition case, but not in the open boundary
case. We will see that $n=n^{\prime }=1$ corresponds to the ground
state. For the limit $c=0$, we have $2\alpha +\beta =3\pi /L$
which means that three atoms all occupy the lowest energy level.
In the other limit $c\rightarrow -\infty $, it is not obvious. By
numerically solving the Bethe ansatz equations, we can get the
values of $\alpha $, $\beta $ and $\Lambda$. Corresponding to
$n=n^{\prime }=1$, we get three sets of solution as shown in Fig.
3. Among the three solutions, we get $ 2\alpha +\beta =\pi /L$
corresponding to the solution $\Lambda =-c+\delta $ (the solution
denoted by squares in Fig. 3), where $ \delta $ is a small number
in the large $c $ limit. This solution is the trimer-like solution
describing the ground state of the attractive 3-atom system.

\end{document}